\documentclass[%
 reprint,
 amsmath,amssymb,
 aps,
]{revtex4-2}

\usepackage{multirow}
\usepackage{verbatim}
\usepackage{graphicx}
\usepackage{dcolumn}
\usepackage{bm}

\begin{document}


\title{Search for Electron Capture of $^{176}$Lu in Lutetium Yttrium
OxyorthoSilicate}

\author{Luigi Ernesto Ghezzer}
\author{Francesco Nozzoli}\email{Francesco.Nozzoli@cern.ch}
\author{Riccardo Nicolaidis}
\author{Roberto Iuppa}
\author{Paolo Zuccon}
\affiliation{%
INFN-TIFPA \& Phys. Dep. Trento University \\
 Via Sommarive 14 I-38123 Trento, It
}%

\author{Cristian De Santis}
\affiliation{INFN-Sezione di Roma Tor Vergata, V. della Ricerca Scientifica 1, I-00133 Rome, Italy}%

\date{\today}

\begin{abstract}
The nuclide $^{176}$Lu is one of the few naturally occurring isotopes 
that are potentially unstable with respect to electron capture (EC).
Although experimental evidence for $^{176}$Lu EC decay is still missing, this isotope is instead well known to  $\beta^-$ decay into $^{176}$Hf with an half-life of about 38 Gyr. The precise investigation of all possible decay modes for $^{176}$Lu is interesting because the Lu/Hf ratio is adopted as an isotopic clock. 
Previous searches for the $^{176}$Lu EC decay were performed by using a passive lutetium source coupled with  an HP-Ge spectrometer. 
Our approach uses a Lutetium Yttrium
OxyorthoSilicate (LYSO) crystal both as lutetium source and as an active detector. Scintillation light from the LYSO crystal is acquired together with the signals from the  HP-Ge detector, this allows a powerful suppression of the background sourcing from  the well known $\beta^-$ decay branch. This approach  led to an improvement  on  the $^{176}$Lu EC branching ratio limits by  a factor 3 to 30, depending on the considered EC channel.     
\end{abstract}

\maketitle


\section{Introduction}
The naturally occurring isotope $^{176}$Lu (2.6\% abundance) it is known to $\beta^-$ decays to $^{176}$Hf with a half-life of $\approx$38 Gyr. 
However, $^{176}$Lu is also one of the six naturally occurring isotopes 
that are potentially unstable with respect to electron capture (EC).
In particular, evidence for EC decay was found for $^{40}$K \cite{TOI}, $^{50}$V \cite{50VNagorny,50Vdanevich} and $^{138}$La \cite{TOI},
but is still missing for $^{123}$Te \cite{123TeCZT}, $^{176}$Lu \cite{Norman2004} and $^{180m}$Ta \cite{180Ta}.
%
The precise investigation of all possible radioactive decay modes of $^{176}$Lu is interesting since the Lu/Hf ratio is an isotopic clock to date meteorites and minerals. 
 In particular, it has been suggested that some discrepancies involving Lu/Hf age comparisons in different samples could be reconciled if $^{176}$Lu also underwent signiﬁcant electron capture (EC) decay \cite{Norman2004,KOSSERT2013140,HULT2014112}.
A second interesting feature is that  $^{176}$Lu/$^{175}$Lu can be considered also as an s-process thermometer in studies of stellar nucleosynthesis \cite{Laeter2006}.

\begin{figure}[h]
\includegraphics[width=0.49\textwidth]{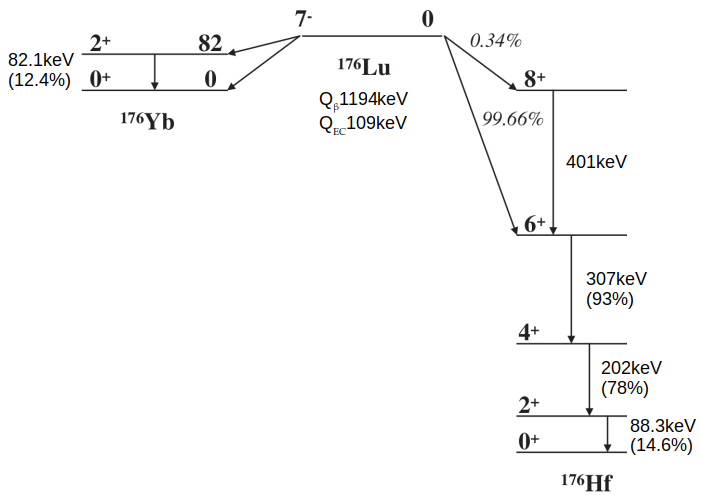}
\caption{\label{fg:Schema} Decay scheme of $^{176}$Lu \cite{NuclearIAEA,QVAL}.              
}
\end{figure}

Figure \ref{fg:Schema}  shows the decay scheme of $^{176}$Lu, on the right side the dominant decay chain to $^{176}$Hf is depicted, while on the left side there is the expected EC decay process to $^{176}$Yb.
  The Q-value for EC decay of $^{176}$Lu (J$^{\pi}$= 7$^-$) to the $^{176}$Yb ground state (J$^{\pi}$= 0$^+$) is $\simeq$ 109 keV \cite{QVAL,Huang_2021,Wang_2021_2},  and the one to the $^{176}$Yb ﬁrst excited state (J$^{\pi}$= 2$^+$) is $\simeq$ 27 keV.
 These EC decay branches, however, would be forbidden transitions of 7$^{th}$ and 5$^{th}$ order respectively. The relative branching ratios are then  expected to be very small, also considering that  forbidden transitions of order  $>$4$^{th}$ have  never been experimentally observed so far \cite{logft1998}.

Previous searches for the $^{176}$Lu EC decay were performed by using a passive lutetium source coupled to a HP-Ge detector  use to search for the $^{176}$Yb$^*$ 82 keV $\gamma$-ray or the characteristic Yb x-rays
\cite{Norman2004}.
Our approach, that we call ``active source'', uses a Lutetium–Yttrium OxyorthoSilicate (LYSO \cite{LYSOSG}) crystal scintillator coupled to a PMT  both as a $^{176}$Lu source and as detector. The LYSO crystal PMT signal was acquired together with the signal from an HP-Ge, this allows a powerful reduction of the large background from the $\beta^-$ decay branch. Moreover the possibility of a simultaneous measurement of the energy release within the LYSO crystal allows also the investigation of the different EC decay channels.

In these  cases of highly forbidden transitions, that  are suppressed by a large variation of the nucleus spin as compared with the sum of the angular momentum of the captured electron and of the emitted neutrino, an important channel for the decay is provided by the Radiative Electron Capture (REC) \cite{Detour1992}.
In this process an x-ray can be radiated either by the captured electron (Internal Bremsstrahlung) or by the father/daughter nucleus (detour transitions), this photon contributes with his spin to the overall angular momentum conservation.   
The REC decay has been experimentally observed for the forbidden decay of: $^{41}$Ca, $^{59}$Ni, $^{81}$Kr,$^{137}$La and $^{204}$Tl \cite{REC41Ca,REC59Ni,REC81Kr,REC137La,RECTl204}.

In our setup the energy released by an occurring REC x-ray, combined with  the energy released by the atomic relaxation of the electron vacancy, would be measured by the LYSO scintillator. This measurement when combined with the signals from the  HP-Ge detector, is boosting the  capability to identify  the $^{176}$Lu EC decays.  

\section{Experimental set-up}

\begin{figure}[h]
\includegraphics[width=0.5\textwidth]{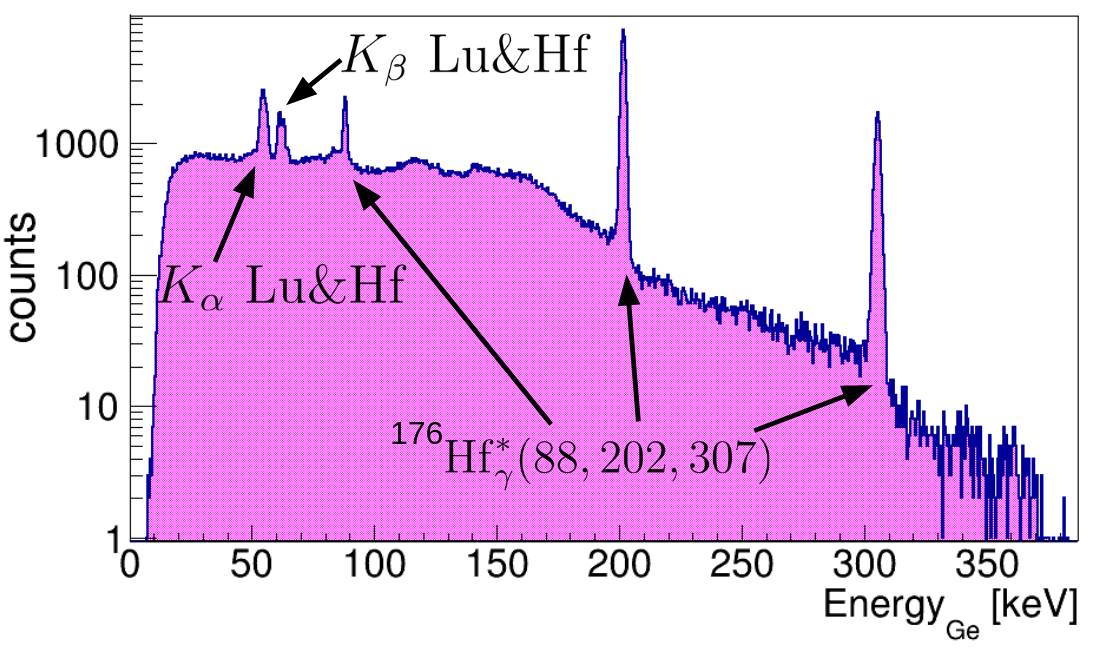}
\caption{\label{fg:HPGeCalib} HP-Ge energy spectrum measured in $\simeq$90h exposure.}
\end{figure}

The measurements were performed using  the HP-Ge facility at the physics department of the Trento University; 
a flat sliced LYSO crystal (elliptical shape of size $\approx$ 35x20x2mm$^3$ - 7.9g) was coupled to an Hamamatsu-R5946 ($\oslash 38$mm) PMT with EJ-550 optical couplant.
The expected LYSO source activity is $\simeq$ 40Bq/g due to the known $^{176}$Lu $\beta$-decay.
This active LYSO source was placed in front of a CANBERRA GC2020 HP-Ge \cite{GC2020} and the two detectors were shielded with 2.5mm of Cu and 5cm of Pb
to minimize the environmental $\gamma$ background. The inner Cu shield is meant to stop the 84.9keV Pb-K$_{\beta}$ x-rays 
that are produced by the external Pb shield and that would provide a background near to the 82.1 keV $^{176}$Yb$^*$ signal region.
Both detector signals were digitized by a LeCroy HDO9104 \cite{HDO9104} in a 20$\mu$s wide time window, the acquisition trigger was set on the HP-Ge signal  with an energy threshold of $\approx 15$keV.
No trigger conditions were imposed to the LYSO signals, this allows to study all the possible EC channels, like the previous experiments using the passive $^{176}$Lu source. A total number of 1.5 Mevents were acquired during 90h exposure time. 
LYSO is a good scintillating material ($\simeq$33ph/keV) with a fast ($\simeq$40ns) decay time, despite the signal of HP-Ge is much slower, 
we achieved a coincidence time resolution of $\simeq$ 100ns in our measurements.
Energy calibration of HP-Ge detector was done using the characteristics $\gamma$-lines produced by $^{176}$Lu $\beta$-decay in $^{176}$Hf$^*$ excited levels.
FIG. \ref{fg:HPGeCalib} shows  the energy spectrum measured by our HP-Ge detector.

\begin{figure}[h]
\includegraphics[width=0.5\textwidth]{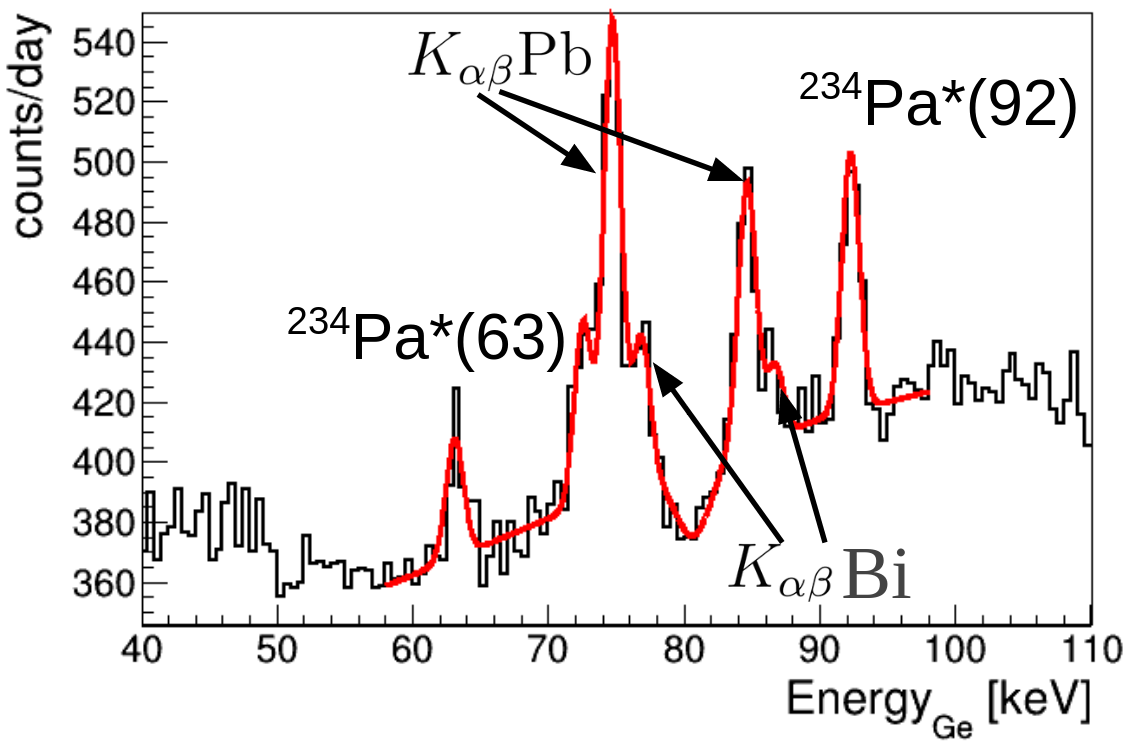}
\caption{\label{fg:HPGe_bkg} Intrinsic/environmental HP-Ge background spectrum. Red line is a background model, as a guide to the eye, where the $\gamma$-lines of $^{234}$Pa$^*$ and x-rays from Pb and Bi are identified.}
\end{figure}

The HP-Ge detector,  shows an energy resolution of 1.50 $\pm$ 0.05 keV FWHM  at 88keV, and 1.65 $\pm$ 0.10 keV FWHM at 55keV.
A measurement of the background for the whole experiment was performed by removing only the thin LYSO crystal from the set-up,
the HP-Ge energy spectrum acquired during a 90h exposure time is shown in FIG. \ref{fg:HPGe_bkg}.
Around 82 keV in the region of interest for the $^{176}$Lu EC, it is possible to identify the residual Pb x-rays: 84.9keV K$_{\beta}$,
75 keV K$_{\alpha1}$ and 72.8 keV K$_{\alpha2}$ surviving the inner Cu shield.
Beyond the expected Pb x-rays, also the characteristics $\gamma$-lines of $^{234}$Pa$^*$ are identified in the background; 
they are occurring in common materials being the $^{234}$Pa a daughter isotope in the $^{238}$U natural radioactive chain \cite{gchron-4-213-2022}. Also an hint for the possible contribution of Bi x-rays to the intrinsic background is observable. In particular $^{214}$Bi and $^{210}$Bi belongs to the $^{238}$U natural radioactive chain whereas $^{212}$Bi belongs to the $^{232}$Th one. 

The large density and effective atomic number (7.1g/cm$^3$, Z$_{eff}$=65) provides a good $\gamma$-ray detection capability for the LYSO scintillator,
at the same time this put a limit to the probability of $\gamma$-ray to escape the crystal and reach the HP-Ge detector. This dictates the thin slice geometry we chose for our LYSO crystal. The overall $\gamma$-ray detection efficiency can be estimated by comparing the measured HP-Ge spectrum with the known intensity ratios of K$_{\beta1,2,3}$/K$_{\alpha1,2}$ lines in Lu and Hf and comparing with the amplitude of $^{176}$Hf$^*$ $\gamma$-lines once
 the internal conversion coefficients are taken into account\cite{NuclearIAEA,TOI,BRICC}.

\begin{figure}[h]
\includegraphics[width=0.49\textwidth]{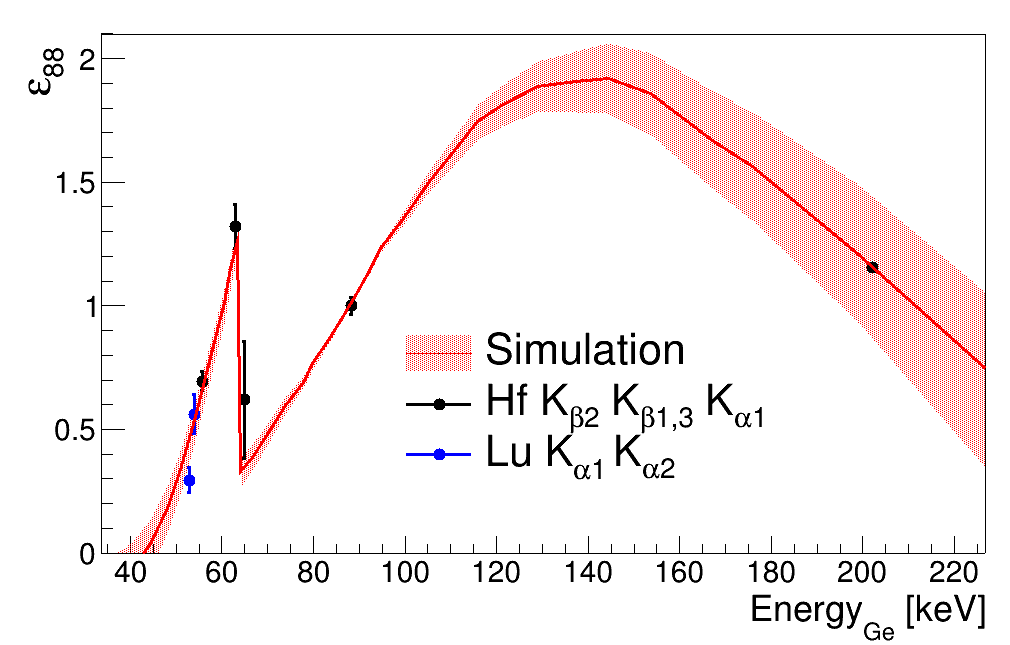}
\caption{\label{fg:effige} HP-Ge detection efficiency, $\varepsilon_{88}$,
(relative to 88keV) for $\gamma$/x-rays emitted by the LYSO active source. 
The red line shows the expected behaviour considering the self-absorption of the LYSO crystal \cite{LYSOSG} combined with the efficiency of GC2020 HP-Ge detector \cite{GC2020}.}
\end{figure}

The measured detection efficiency, $\varepsilon_{88}$, relative to the efficiency for detection of $\gamma$-ray with 88keV energy, is shown in FIG. \ref{fg:effige}. 
The efficiency behaviour is dominated, at low energy, by the photoelectric cross section on lutetium, it is interesting to note that Hf K$_{\beta2}$ x-rays (65keV) are just above the 63.3keV Lu k-edge, whereas Hf K$_{\beta1,3}$ x-rays (63.2keV and 63keV) are just below. The red line in FIG. \ref{fg:effige} shows the expected $\varepsilon_{88}$ behaviour considering the self-absorption of 2mm thick LYSO crystal\cite{LYSOSG} combined with the efficiency of GC2020 HP-Ge detector \cite{GC2020}.
The efficiency for $^{176}$Yb$^*$ 82.1keV is $\simeq$80\% with respect to the one of 88keV, 
whereas the efficiency for Yb K$_{\beta}$ (59.3keV) is almost the same. 
%

\begin{figure}[h]
\includegraphics[width=0.43\textwidth]{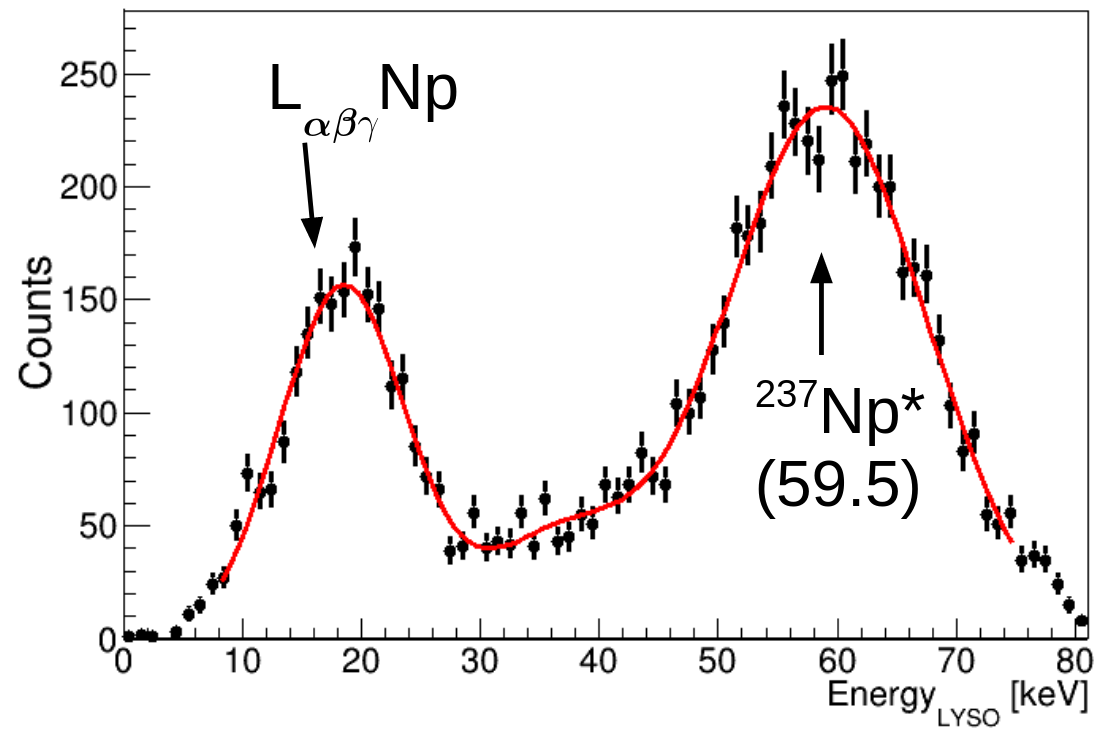}
\caption{\label{fg:CalLYSO} LYSO scintillator energy spectrum for $^{241}$Am calibration source. Red line is a multi-gaussian fit considering the 59.5keV peak from the $^{237}$Np$^*$ decay and the folded contribution of L$_{\alpha\beta\gamma}$ Np x-rays to measure detector energy-resolution.}
\end{figure}

The calibration of the LYSO scintillator was performed with  $^{241}$Am and $^{137}$Cs external sources.
 FIG. \ref{fg:CalLYSO} shows the collected energy spectrum using $^{241}$Am; beyond the known 59.5keV peak from the $^{237}$Np$^*$ decay, it is also possible to observe, at lower energy, the folded contribution of 13.9-17.8-20.8 keV due to L$_{\alpha\beta\gamma}$ Np x-rays.

\begin{figure}[h]
\includegraphics[width=0.49\textwidth]{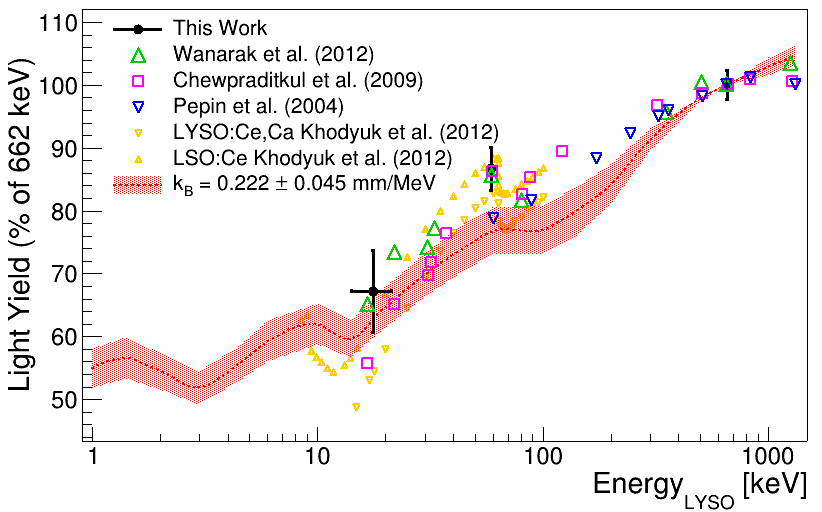}
\caption{\label{fg:birks} The light yield measured for our LYSO crystal (black points) is compared with the existing published measurements for other similar crystals \cite{Pepin2004,Chewpraditkul2009,WANARAK2012,Khodyuk2012}.  
The red line is the result of a Geant4 simulation of the expected light yield using the quenching parameters of ref. \cite{Adriani2022}
and varying $k_B$ within the measurement uncertainty (red line/area).}
\end{figure}

LYSO scintillators are affected by a non-proportionality for $\gamma$-ray response at low energy \cite{Pepin2004,Chewpraditkul2009,WANARAK2012}.
Light yield for x-rays in the keV region drops to $\approx$ half of the one in the MeV region. A few \% of variations of light yields measured among different crystals is possibly due to the different Y and Ce concentrations \cite{Khodyuk2012}.
This non-proportionality effect is due to the scintillation quenching caused by the high ionization density of the relatively slow electrons produced in the low-energy x-ray conversion.
A recent measurement of Birks-Onsager quenching parameters for LYSO scintillator used a 30 GeV/n Argon beam (and nuclear fragments) \cite{Adriani2022}. There the luminous efficiency of LYSO was modeled as:
\begin{equation*}
	L_{eff}=\left(1-\eta_{e/h} e^{-k_o\frac{dE}{dx}}\right)  \left( \frac{1-\eta_H}{1+(1-\eta_H)k_B\frac{dE}{dx}}+\eta_H\right)
\end{equation*}
where the first factor describes the Onsager mechanism while the second one accounts for the (modified) Birks' law \cite{Adriani2022}.  
In figure \ref{fg:birks} the light yield measured for our LYSO crystal (black points) is compared with the existing published measurements for other similar crystals \cite{Pepin2004,Chewpraditkul2009,WANARAK2012,Khodyuk2012}.  
A Geant4 \cite{GEANT} simulation of the expected light yield using the quenching parameters of ref. \cite{Adriani2022} is also shown for comparison (red line). In this work the energy scale of our LYSO scintillator has been evaluated taking into account the light yield non-proportionality expected from this simulation, we check that different modeling of the light yield non-proportionality gives a small/negligible impact in the results.

%

\begin{figure}[h]
\includegraphics[width=0.49\textwidth]{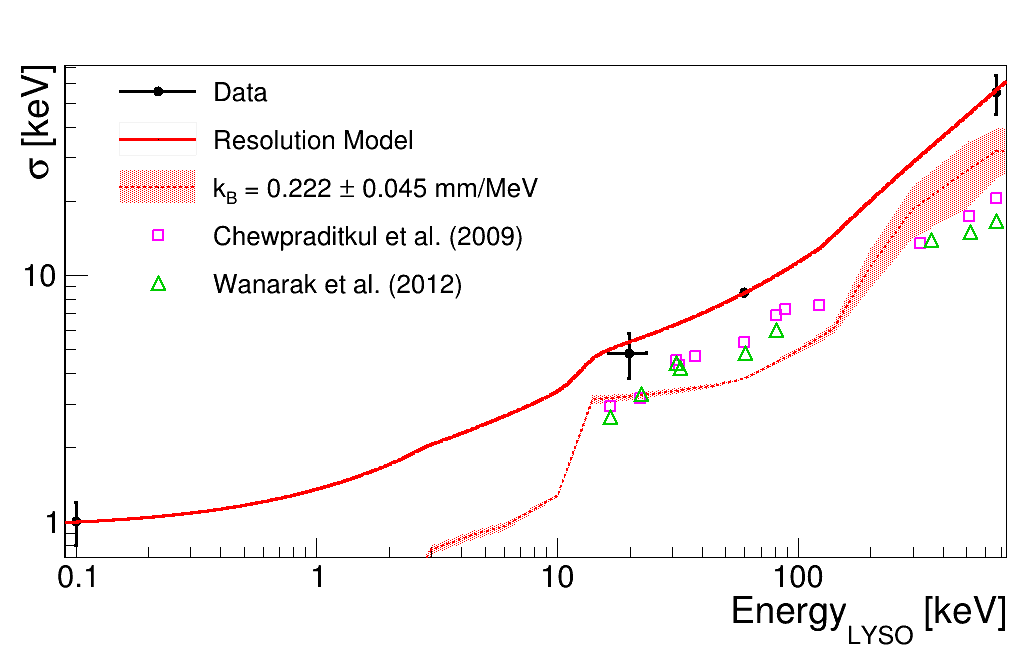}
\caption{\label{fg:ResLYSO} Measured energy resolution for the LYSO scintillator (black points). The red dashed line is the expected LYSO intrinsic energy resolution, $\sigma_I$, evaluated with a Geant4 simulation using the quenching parameters of ref. \cite{Adriani2022}
and varying $k_B$ within the measurement uncertainty (filled area). The intrinsic resolution published for similar LYSO scintillator is shown for comparison (squares and triangles \cite{Chewpraditkul2009,WANARAK2012}). The red continuous line is the model of the energy resolution measured for our LYSO scintillator.}
\end{figure}

The measured energy resolution of the LYSO scintillator is shown in fig. \ref{fg:ResLYSO}. It is important to note that for the LYSO scintillator, an intrinsic energy resolution, $\sigma_{I}$, exists. This is due to the different processes involved in the $\gamma$-ray conversion providing electrons of different energies (thus differently quenched).
The intrinsic resolution evaluated by a Geant4 simulation 
using the quenching parameters of ref. \cite{Adriani2022} is also shown in fig. \ref{fg:ResLYSO} and compared with the LYSO intrinsic resolution measured by \cite{Chewpraditkul2009,WANARAK2012}.
The energy resolution of our LYSO scintillator was modeled as:
$\sigma_{LYSO}=\sqrt{\sigma_0^2+E/n^*+\sigma_{I}^2}$
where $\sigma_0\simeq 1$keV is the electronic noise contribution (the LYSO pedestal was measured with $^{241}$Am source on the HP-Ge), $n^* =1.8 \pm 0.5$ ph.e/keV is the number of collected photoelectrons for detected energy unity and $\sigma_I$ is the expected intrinsic resolution for the LYSO.    

\section{Search for $^{176}$Lu EC}

\begin{figure}[h]
\includegraphics[width=0.49\textwidth]{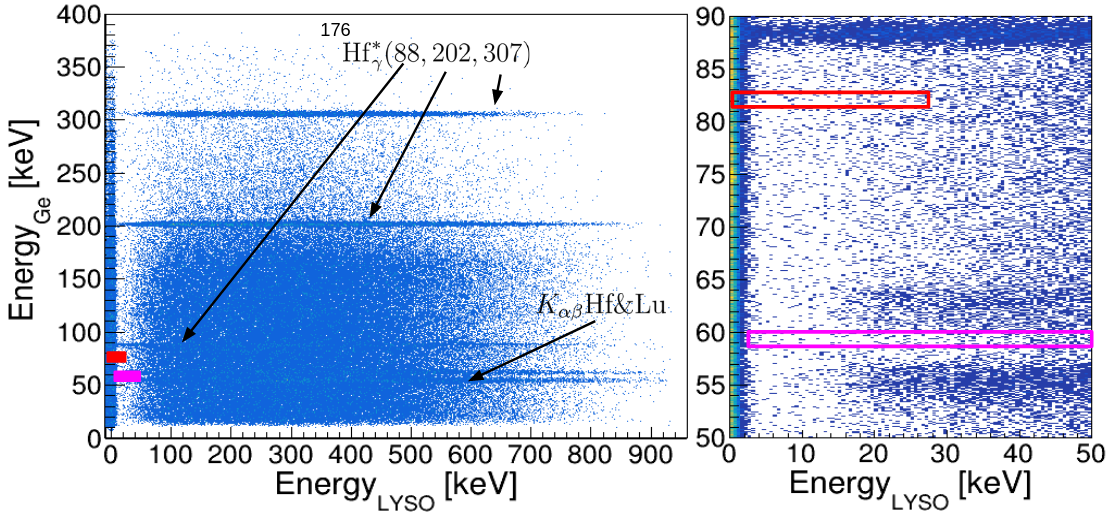}
\caption{\label{fg:spectra2d} Energy$_\mathrm{Ge}$ vs Energy$_\mathrm{LYSO}$ spectrum and a zoom in the region of interest for EC. Red and magenta boxes are the region of interest for $^{176}$Lu EC
to $^{176}$Yb$^*$ and to ground states, respectively.}
\end{figure}

Most of the events collected by the HP-Ge detector come from the $^{176}$Lu $\beta$-decay and the intrinsic/environmental background.
These two classes of events are well recognized in the bidimensional spectrum of FIG. \ref{fg:spectra2d}.
In particular, the population of events characterized by Energy$_\mathrm{LYSO}<$3keV (vertical strip in FIG. \ref{fg:spectra2d}) are mostly due to
intrinsic or environmental background.
On the other hand, events characterized by Energy$_\mathrm{LYSO}>$3keV are mostly due to the continuous electron distribution emitted by $^{176}$Lu $\beta$-decay.
For this last class of events, in FIG. \ref{fg:spectra2d} are also visible the $^{176}$Hf$^*$ $\gamma$-lines characterized by a fixed energy detected in the HP-Ge and the relative continuous distributions due to the Compton scattering. 
Similarly, at lower energy, the Hf x-rays due to internal conversion
of $^{176}$Hf$^*$ levels and the Lu x-rays due to photoelectric absorption, within the
crystal, of $^{176}$Hf$^*$ $\gamma$-rays, are visible.

The powerful background rejection provided by the active source technique can be noticed in FIG. \ref{fg:spectra2d}. All the events with LYSO energy larger than 27keV or 50keV are rejected and the region of interest for the $^{176}$Lu EC decay to $^{176}$Yb excited and ground state shrinks down to the red and magenta boxes.
The measurement of the energy deposited in the LYSO allows to tag (and reject) most of the background from the $^{176}$Lu $\beta$-decay. 

\subsection{Search for $^{176}$Lu EC to $^{176}$Yb$^*$}
Considering the 5$^{th}$ forbidden EC transition of $^{176}$Lu to $^{176}$Yb$^*$, a strong identification signature is the $82.1$keV de-excitation $\gamma$-ray observed in the HP-Ge detector.
The maximum neutrino energy for this decay is:
\begin{equation*}
E^{\nu_{max}}_{82}=Q_{EC}-E_b-82.1 keV
\end{equation*}
where E$_b$ is the Yb atomic binding energy of the captured electron \cite{binding}, therefore the capture of the
\textit{1s} electron is not energetically allowed for this detection channel.

\begin{figure}[h]
\includegraphics[width=0.45\textwidth]{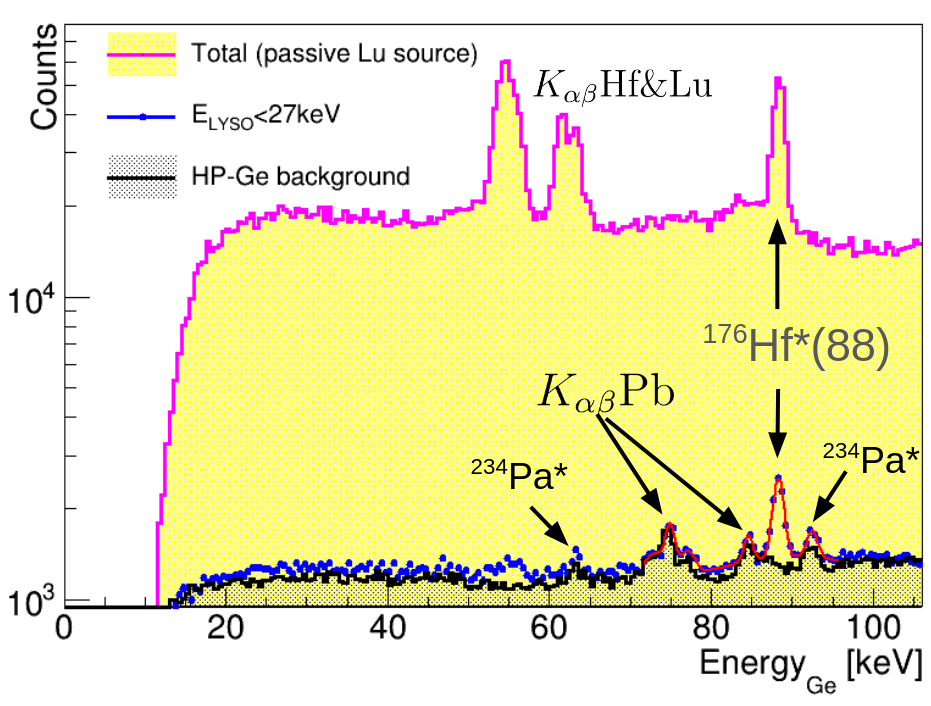}
\includegraphics[width=0.45\textwidth]{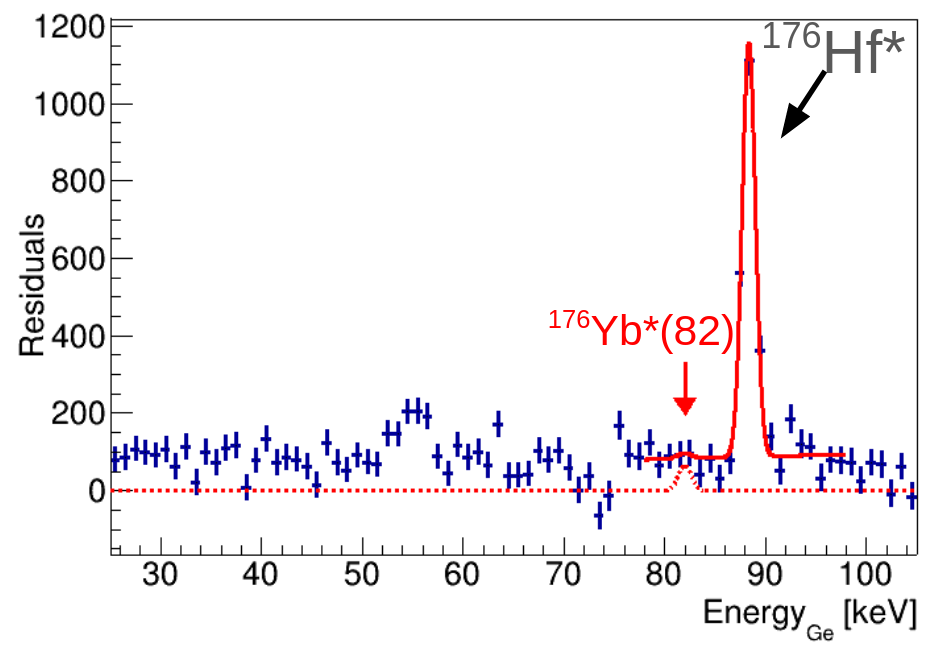}
\caption{\label{fg:res82} Search for the $^{176}$Yb$^*$ $82.1$keV de-excitation $\gamma$-ray. The cut E$_\mathrm{LYSO}<$27keV allows
a background reduction of a factor $\approx$ 20 with respect to the "passive source" approach. The red arrow points out the expected position for the $^{176}$Yb$^*$ $82.1$keV peak in the background subtracted residual spectrum.
The red continuous line is a model considering Gaussian peaks over a flat background, the red dashed line is the 95\% upper limit for the 82.1keV peak contribution.}
\end{figure}

In the top panel of FIG. \ref{fg:res82} the energy spectrum measured with our HP-Ge detector is shown (magenta line). These data are plotted without any requirement on  the LYSO measured energy,  and as a consequence this plot is dominated by the background from the $^{176}$Lu $\beta$-decay. This corresponds to  typical landscape  of  the passive source measurement approach used in \cite{Norman2004}. 
   
Profiting of the "active source" technique, most of $^{176}$Lu $\beta$-decay events are rejected by the request E$_\mathrm{LYSO}<$27keV, i.e. the maximum expected energy in LYSO for the EC (blue points on FIG. \ref{fg:res82} top panel).

This approach provides a factor $\approx$20 in background suppression, and the resulting measured spectrum mostly corresponds to  the intrinsic/environmental background characteristic of our HP-Ge detector. The black points on FIG. \ref{fg:res82} top panel, are indeed our measurement of such background already shown in FIG. \ref{fg:HPGe_bkg}. The other relevant feature is the 88.3 keV peak from the $^{176}$Hf* last de-excitation, that becomes evident observing the residuals after the HP-Ge background subtraction, see  FIG. \ref{fg:res82} lower panel. N$_{82}$=38$\pm$105 events are obtained fitting the residuals, thus no statistical evidence for the $^{176}$Yb$^*$ 82.1keV peak was found. The  corresponding 95\% C.L.  upper limit for a  $^{176}$Lu EC process  is estimated to 210 events  (dotted line in FIG. \ref{fg:res82} lower panel).

Following the approach of \cite{Norman2004} the upper limit on the EC branching fraction can be obtained by comparing the upper limit on the number
of the $^{176}$Yb$^*$ $82.1$keV de-excitation $\gamma$-rays with the number of $^{176}$Hf$^*$ events measured by fitting the 88.3keV $\gamma$-line in the whole event distribution, thus related to the $\beta$-decay branching fraction
(N$_{88}$=119.9 $\pm$ 0.5 $\times$ 10$^3$ events):
\begin{equation}
B_{82}=\frac{(1+\alpha_T^{82})N_{82}/\varepsilon_{88}}{(1+\alpha_T^{88})N_{88}}<2.6\times 10^{-3} \; \mathrm{(95\%C.L.)} 
\end{equation}
where $\alpha_T^{82}$=7.06 and $\alpha_T^{88}$=5.86 are the total electron conversion coefficients for first excited levels of $^{176}$Yb and $^{176}$Hf, respectively \cite{NuclearIAEA,BRICC}, and $\varepsilon_{88}(82)\simeq 80\%$ is the detection efficiency loss of 82keV $\gamma$-ray with respect to 88keV $\gamma$-ray.

\subsection{Search for REC or specific EC channels}

The  result presented in the previous section represents a cautious upper limit that holds for all possible channel of EC  of $^{176}$Lu to $^{176}$Yb$^*$.
 However, with our setup, a specific energy distribution is expected to be observed with  the LYSO crystal, for each kind of EC process.
In particular, depending on the specific (sub-)shell of the captured Lu electron, the relative binding energy is released in the LYSO crystal as a consequence of the Yb atomic rearrangement, emitting low energy x-rays or Auger electrons.  
\begin{figure}[h]
\includegraphics[width=0.49\textwidth]{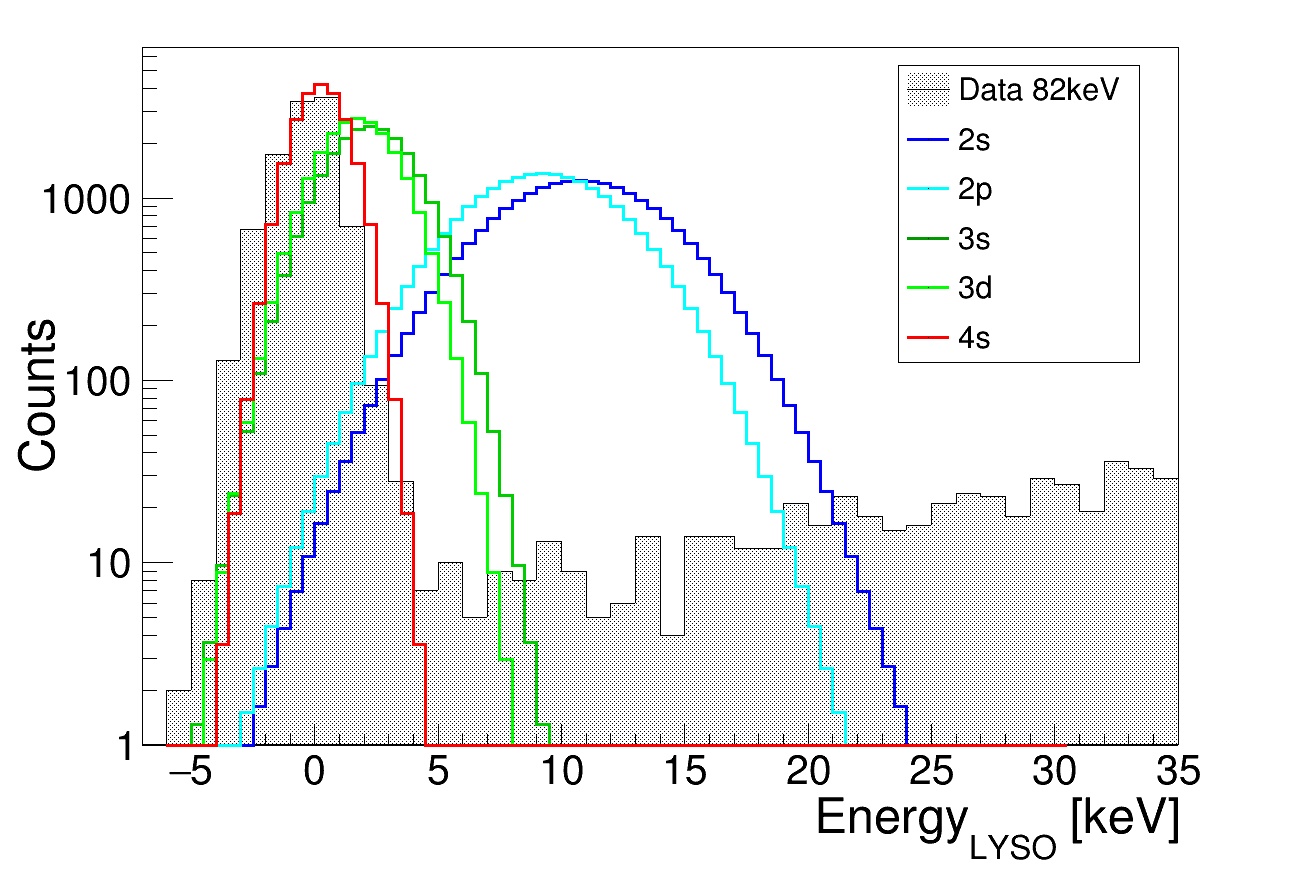}
\caption{\label{fg:EC82s} Expected energy distributions in the LYSO crystal due to L-M-N shell electron captures.
The shaded energy distribution is the measured one by selecting 82$\pm$2keV energy deposited in the HP-Ge detector.}
\end{figure}

FIG. \ref{fg:EC82s} reports the expected energy releases in the LYSO crystal from  L-M-N shell electron captures. These distributions are compared with the measured energy distribution in the LYSO scintillator for those events where the energy detected in HP-Ge detector is 82$\pm$2keV (shaded area). The measured signal is characterised by a peak at $E_{LYSO}=0$, due to intrinsic/environmental background, and by a continuous distribution due to the electron energy emitted by the  $^{176}$Lu $\beta$-decay.  

On the other hand, in case of a Radiative Electron Capture, also the low energy x-ray emitted is detected in the LYSO crystal together with the electron binding energy.
Following the formalism of \cite{REC1956}, the x-ray energy distribution expected for a REC is:
\begin{equation}
\frac{dN_x}{dk} (k) = N_0 R_x(k) k \left(1-\frac{k}{q_x}\right)^2
\end{equation}
where k is the x-ray energy, x denotes the considered atomic electron, $q_x$ is the maximum allowed x-ray energy in the REC, $R_x(k)$ is a model dependent shape factor and N$_0$ is an overall normalization.

\begin{figure}[h]
\includegraphics[width=0.49\textwidth]{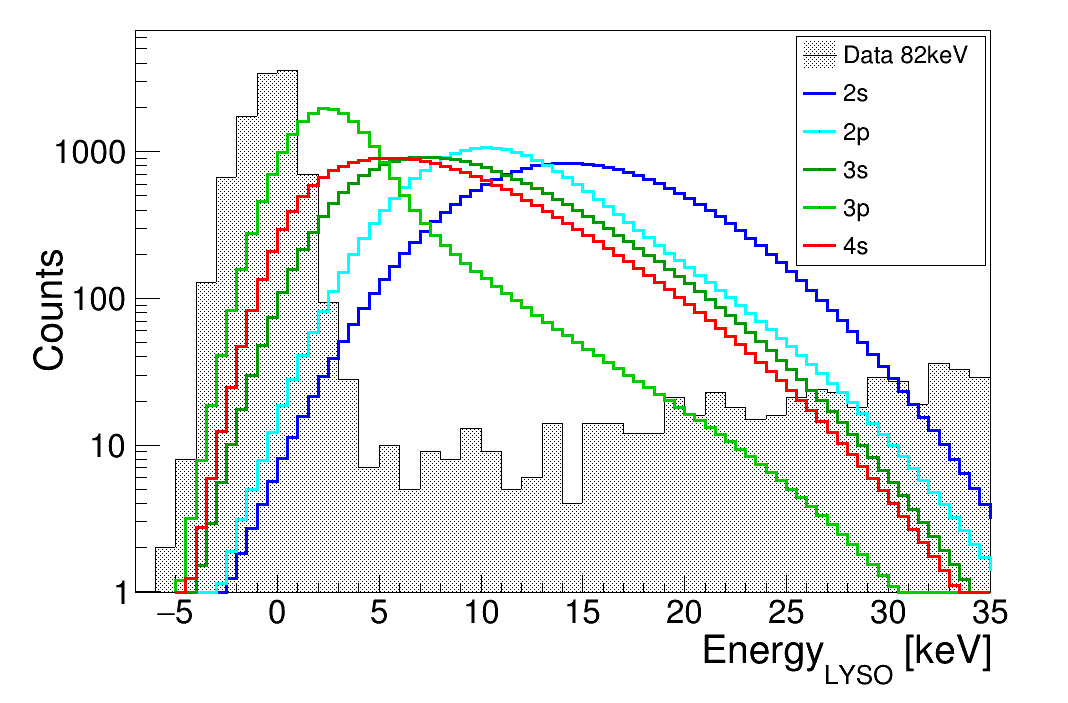}
\caption{\label{fg:REC82s} Expected energy distributions in the LYSO crystal due to radiative electron captures.
The shaded energy distribution is the measured one by selecting 82$\pm$2keV energy deposited in the HP-Ge detector.}
\end{figure}

For the Internal Bremsstrahlung of s-state electrons (characterized by a larger wavefunction overlap with the nucleus) an analytic model for $R_{ns}$ can be produced following some reasonable approximations\cite{REC1956}. However, the general analytical form for the shape factor of REC of \textit{ns} level electrons can be expressed as \cite{REC137La}:
\begin{equation}
\begin{array}{c}
R_x(k) = A_x^{(1)}\left(1-\frac{k}{q_x}\right)^2 + A_x^{(2)}\Lambda \left(\frac{k}{q_x}\right)^2 + \\ 
+ 2 \left(\frac{e_\mathrm{eff}m_e}{q_x}\right)^2 + A_x^{(3)} e_\mathrm{eff} \frac{m_e k}{q^2_x}
\end{array}
\end{equation}

\noindent where $A_x^{(i)}(k)\approx 1$ are describing Coulomb effects, $\Lambda$ represents a combination of reduced nuclear matrix elements and the last two terms arise from  the possible contribution due to Detour Transitions (DT). 
In the latter case,  the parameter $e_\mathrm{eff}$, the effective charge, represents the strength of the DT.
Entering into the  details of a  model of the possible REC for $^{176}$Lu is beyond the purpose of this work, however it is interesting to note that for the case of the measured REC of heavy nuclei, like $^{137}$La and $^{204}$Tl, the contribution of Detour Transitions appears to be negligible \cite{REC137La,RECTl204}. Moreover we also verify that the variation of $\Lambda$ from 0 to 1 introduces only small modifications (within 10\%) of our results. We fix, for simplicity, $\Lambda$=0 and $e_\mathrm{eff}$=0, thus assuming the minimal contribution of REC x-ray energy in the LYSO spectrum.
Regarding REC from 2p and 3p shells we derive the shape factors from the numerical tables in \cite{REC1956}.
In figure \ref{fg:REC82s} the energy distributions expected in the LYSO scintillator for some REC channels are shown.

\begin{figure}[h]
\includegraphics[width=0.49\textwidth]{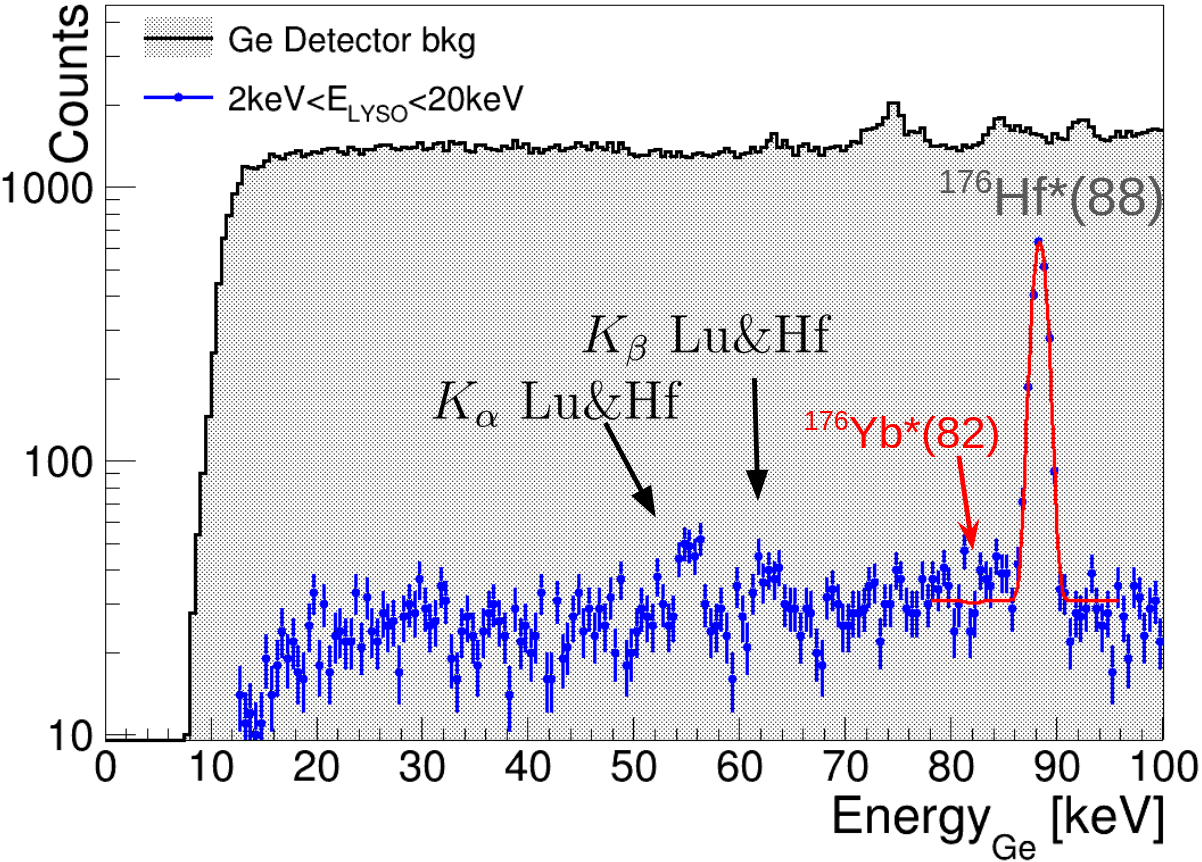}
\caption{\label{fg:res_2_20} Search for L-shell EC/REC or for ns-shell REC of $^{176}$Lu in the $^{176}$Yb$^*$ $82.1$keV level. The cut 2keV$<E_\mathrm{LYSO}<$20keV allows further background reduction of a factor $\approx$ 40 with respect to the intrinsic/external detector background (shaded). Red line is a fit model considering the 88.3keV $^{176}$Hf$^*$ and the possible 82.1keV $^{176}$Yb$^*$ peaks over a flat background.}
\end{figure}

The analysis proceeds by searching for the 82.1 keV peak in the HP-Ge energy distributions considering two different selections for E$_\mathrm{LYSO}$ and evaluating the relative signal efficiency from the expected energy distributions of FIG. \ref{fg:EC82s} and \ref{fg:REC82s}.
The selections on E$_\mathrm{LYSO}$ will improve the overall signal over background ratio.
In FIG. \ref{fg:res_2_20} the HP-Ge energy distribution considering the selection 2keV$<E_\mathrm{LYSO}<$20keV is shown,
this allows the rejection of most of the intrinsic detector background providing a further $\approx$40 reduction factor with respect to the counting rate shown in FIG. \ref{fg:res82}.
The number of events that can be attributed to $^{176}$Yb$^*$ is N$_{82}$=-2$\pm$12, thus no statistical evidence for the EC/REC is found in this energy window. Considering the slightly different E$_{LYSO}$ selection efficiencies for the different sub-channels (98\% to 82\%)  the limits to branching ratio: $<0.024\%$ and $\lesssim 0.028\%$ at 95\% C.L. can be inferred for the L-shell EC and n-s REC respectively. Limits on these channels are relative to electrons whose wavefunctions are more overlapped with the nucleus, they are more than one order of magnitude lower when compared with previous ones based on the passive lutetium source \cite{Norman2004}.

\begin{figure}[h]
\includegraphics[width=0.49\textwidth]{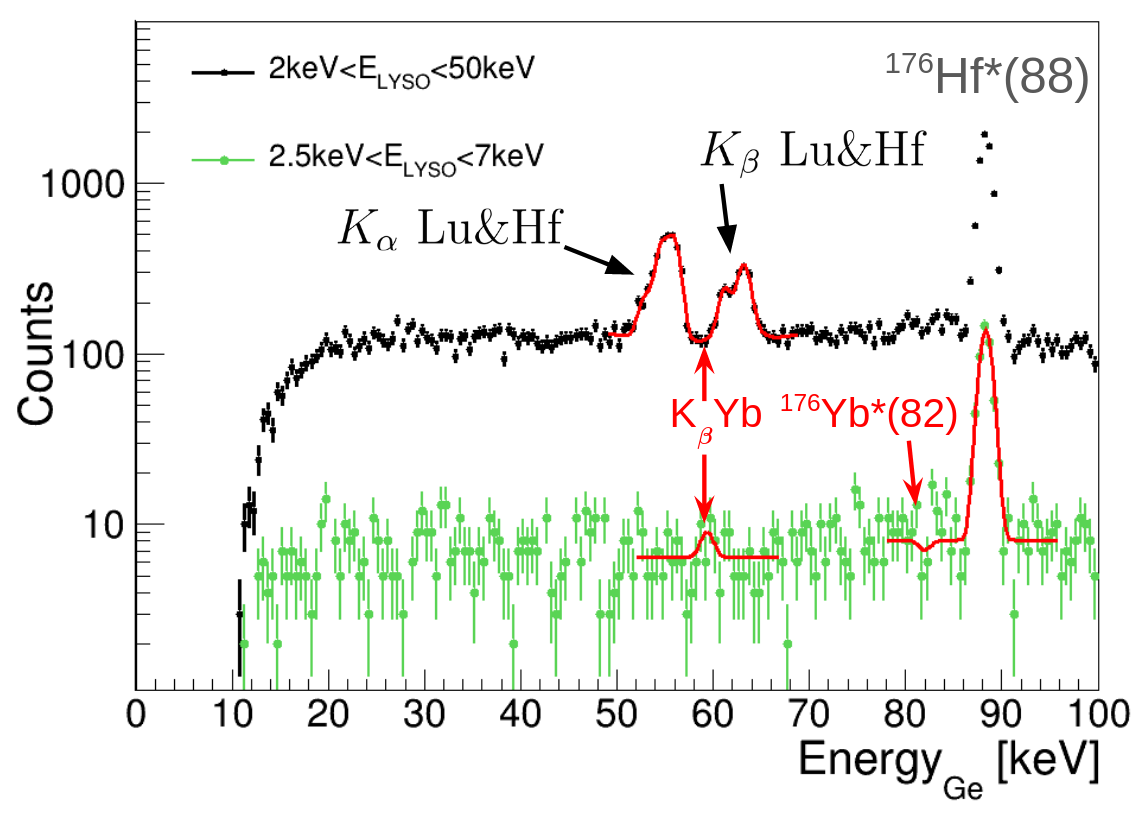}
\caption{\label{fg:res_59} Search for EC/REC of $^{176}$Lu in the $^{176}$Yb$^*$ $82.1$keV level and in the $^{176}$Yb fundamental state. The selection  2.5keV$<E_\mathrm{LYSO}<$7keV allows to search for K-shell EC (K$_\beta$=59.3keV x-ray)  or M/N-shell EC/REC to $82.1$keV level.
The red lines are the fit models to investigate the possible presence of the Yb 59.3keV  K$_\beta$ or the $^{176}$Yb$^*$ $82.1$keV peaks considering the contribution of the other $\gamma$/x-rays Gaussian peaks over a continuous background.}
\end{figure}

On the other hand, FIG. \ref{fg:res_59} shows the measured events in the 2.5keV$<E_\mathrm{LYSO}<$7keV energy window,
this is suitable for the study of the M/N-shell REC channels or M-shell EC channel releasing a smaller energy contribution within the LYSO crystal.
In this case the number of events that can be attributed to $^{176}$Yb$^*$ is N$_{82}$=-3$\pm$6, thus the upper limit to N$_{82}$ is more stringent, however  the selection efficiency for these channels is relatively small (in the $30-45\%$ range).   
Table \ref{tb:table} summarizes the branching ratios limits obtained in the different sub-channels.

Finally, the precise shape factors for REC of d and f electron orbitals are not quantitatively evaluated in literature, however it is reasonable to expect a very soft Internal Bremsstrahlung spectrum for 4f electron REC, providing a negligible energy deposition within the LYSO scintillator. It is important to note that REC of 4f electrons in $^{176}$Lu is a very interesting channel since it allows a neutrino emission with $\hbar$/2 total angular momentum. A dedicated investigation of the region E$_\mathrm{LYSO}<$4keV provides the largest excess, N$_{82}$=167$\pm$100, observed in this experiment, however this is still fully compatible with the expected statistical fluctuations.

\subsection{Search for $^{176}$Lu K-shell EC/REC}
The EC decay of $^{176}$Lu to the fundamental state of $^{176}$Yb is a 7$^{th}$ degree forbidden transition, thus it is expected to be suppressed with respect to the 5$^{th}$ degree forbidden EC decay in $^{176}$Yb$^*$; however if a $^{176}$Lu nucleus captures  a K-shell electron, the result must be a $^{176}$Yb in the ground state. 
Considering that the K-shell atomic orbital is the one with the largest superposition with the nucleus, this possibility deserves a dedicated investigation.

A signature of the K-shell capture is provided by the x-rays emitted when the Yb  atomic vacancy is filled. In particular, K$_{\alpha2}$=51.35keV,  K$_{\alpha1}$=52.4keV and K$_{\beta}$=59.3keV are be emitted with a probability of  
27.2\%, 48.1\% and 15.1\%, respectively \cite{TOI}.
By comparing these x-ray energies with the measured energy distribution (see e.g. FIG. \ref{fg:res_59})
it is clear that the Yb K$_{\beta}$ line is the only one distant enough from the nearby Lu and Hf K-shell lines, to be easily identified. Moreover, due to self absorption in the LYSO crystal and in the HP-Ge dead layers, the expected detection efficiency for Yb K$_{\alpha}$ lines is quite small, whereas the Yb K$_{\beta}$ line has  $\varepsilon_{88}\simeq$100\% 
(see FIG. \ref{fg:effige}).

To search of the Yb K$_{\beta}$ line, we must first recall that the Q-value for $^{176}$Lu EC  is $\simeq$ 109 keV, if the emitted x-ray accounts for 59.3keV, the maximum energy that we might detect in the LYSO crystal is $\simeq$ 50 keV. This value, however, 
 represents only the end point of the possible REC energy distribution.
On the other hand, knowing that the K-shell ionization energy in Yb is 61.3keV, it is possible to estimate a minimum energy of 2keV detected in the LYSO from  the additional x-rays/Auger electrons emitted by the Yb atomic de-excitation.
\begin{figure}[h]
\includegraphics[width=0.49\textwidth]{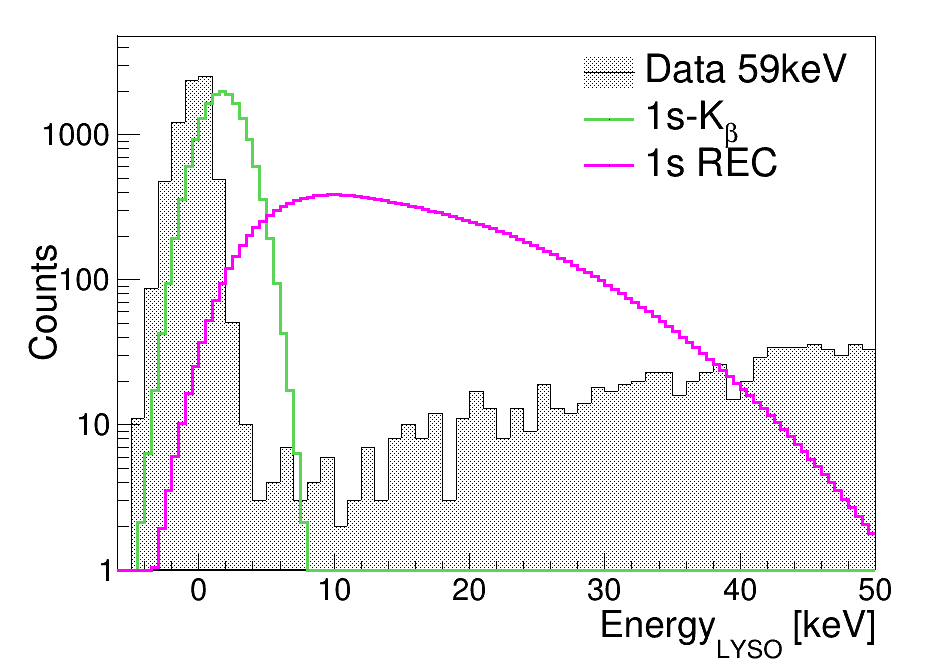}
\caption{\label{fg:sha_59} Expected LYSO energy distributions for EC (green line) and REC (magenta line) of $^{176}$Lu in $^{176}$Yb ground state. As a comparison the LYSO energy distribution for 59.3keV measured in the HP-Ge is shown (shaded area).}
\end{figure}
Assuming a detection of Yb K$_{\beta}$ x-ray in HP-Ge, in FIG. \ref{fg:sha_59}, we show the REC energy distribution (magenta) and the 2 keV contribution folded with the LYSO energy resolution (green). In FIG. \ref{fg:sha_59} the expected EC signal distributions are compared with the  measured LYSO energy distribution for the events where a  59keV signal is observed in the HP-Ge (shaded area).
Similarly to FIG. \ref{fg:EC82s} and \ref{fg:REC82s}, also 
 this measured distribution is characterized by a peak at $E_\mathrm{LYSO}$=0 due to intrinsic/external HP-Ge background and a continuous distribution due to the known $^{176}$Lu $\beta$-decay. 
 
The  measured HP-Ge energy distribution obtained when applying the selection  2keV$<E_\mathrm{LYSO}<$50keV is shown in FIG. \ref{fg:res_59}.
 In this plot the number of events attributed to a possible  Yb K$_{\beta}$ peak are N$_{59}$=3$\pm$42, thus no evidence for REC capture of K-shell electron is found. The upper limit to the corresponding branching fraction is:
\begin{equation}
B_{59}=\frac{N_{59}/(\varepsilon_{59} \varepsilon_{88} \varepsilon_{sel})}{(1+\alpha_T^{88})N_{88}}<6\times 10^{-4} \; \mathrm{(95\%C.L.)} 
\end{equation}
where $\varepsilon_{59}$=15.1\% is the probability of a K$_{\beta}$ x-ray emission filling an Yb K-shell vacancy, 
$\varepsilon_{88}\simeq$ 100\% 
is the detection efficiency of 59keV $\gamma$-ray with respect to 88keV $\gamma$-ray and $\varepsilon_{sel}\simeq$ 98\% is the efficiency for the expected REC distribution considering the 
2keV$<E_\mathrm{LYSO}<$50keV selection.

Restricting the LYSO window to the 2-20keV range ($\varepsilon_{sel}\simeq$ 70\%) the $\beta$-decay background is reduced by a factor $\approx$5, we found N$_{59}$=-2$\pm$15 thus the upper limits to the REC branching fraction improves to $B_{59}<2.9 \times 10^{-4}$ (95\%C.L.).

The pure K-shell EC (green line in FIG. \ref{fg:sha_59}) was investigated by considering the 2.5keV$<E_\mathrm{LYSO}<$7keV selection ($\varepsilon_{sel}\simeq$ 30\%), the number of events attributed  
to a possible  Yb K$_{\beta}$ peak are N$_{59}$=4$\pm$6, thus also for this process no evidence is found and the upper limit  is: $B_{59}<3.5 \times 10^{-4}$ (95\%C.L.).

\section{Discussion and conclusions}

The search for Electron Capture in $^{176}$Lu is an opportunity for a laboratory measurement of  5$^{th}$ degree forbidden processes that may provide  valuable information for nuclear theory models.
Our active source technique, that uses a LYSO crystal  scintillator also as detector, allows a powerful reduction of the background from $^{176}$Lu $\beta$-decay and the HP-Ge intrinsic/external contamination.
No evidence for the EC process was found. Depending on the particular EC channel, we were able to set upper limits to the $^{176}$Lu EC branching fraction that are a factor 3-30 better than what obtained with previous measurements \cite{Norman2004}.
Table \ref{tb:table} reports  our estimation of the upper limits for the different EC channels.

Considering the known $\approx$ 38Gyr half-life of $^{176}$Lu, the obtained limits on the partial half-life for the EC processes are in the range of few $\times$ 10$^{13}$-10$^{14}$y. It is interesting to compare these limits with the partial half-life of the other 5 naturally occurring isotopes that can decay via the EC process:

$^{40}$K (4$^-$) $\rightarrow$ $^{40}$Ar (2$^+$)   $T^{EC}_{1/2}$=1.2$\times$10$^{10}$y \cite{TOI}

$^{50}$V (6$^+$) $\rightarrow$ $^{50}$Ti (2$^+$)   $T^{EC}_{1/2}$=2.7$\times$10$^{17}$y\cite{50VNagorny,50Vdanevich} 

$^{138}$La (5$^+$) $\rightarrow$ $^{138}$Ba (2$^+$)   $T^{EC}_{1/2}$=1.6$\times$10$^{11}$y
\cite{TOI}

$^{123}$Te (1/2$^+$) $\rightarrow$ $^{123}$Sb (7/2$^+$)  $T^{EC}_{1/2}>$3$\times$10$^{16}$y \cite{123TeCZT}

$^{180m}$Ta (9$^-$) $\rightarrow$ $^{180}$Hf (6$^+$)  $T^{EC}_{1/2}>$2$\times$10$^{17}$y \cite{180Ta}.

This comparison suggests that improvements in sensitivity of 2-3 order of magnitude could be possible, in principle, by adopting the techniques
of low-background experiments.
In particular, moving to the HP-Ge facility of an underground laboratory
the intrinsic/external background would be reduced by few order of magnitude.
Moving to an underground laboratory would be important also to suppress the tricky background due to the possible environmental neutron capture
of the (97.4\%) abundant $^{175}$Lu isotope that could produce the $^{176m}$Lu (1$^-$) 123keV level.
It is known that this nuclear isomer, whose half life is 3.6h, undergoes EC populating both the ground state and the (2$^+$) levels of $^{176}$Yb \cite{NuclearIAEA}.
Finally, both the HP-Ge and LYSO energy resolutions can be  improved. In particular
by considering the possibility to operate small LYSO crystals as cryogenic bolometers, a very good signature of EC peaks in the LYSO spectrum is expected.

\begin{table}[h]
  \begin{center}
    \caption{Upper limits on the branching fraction for specific Electron capture channels of $^{176}$Lu.}
    \label{tb:table}
    \begin{tabular}{ p{2.9cm} | p{2.8cm} | p{2.7cm} } 
      \textbf{Electron Capture channel} & \textbf{Branching ratio Limit (95\% C.L.)} & \textbf{Previous Limit (68\% C.L.) \cite{Norman2004}}\\
      \hline
     1s EC + 59keV & \hfil 0.035\% & \hfil \multirow{2}{*}{0.36\%} \\
     1s REC + 59keV & \hfil 0.029\% &  \\
     \hline
     L EC + 82keV & \hfil 0.024\% & \hfil \multirow{8}{*}{0.45\%} \\
L/3s REC + 82keV & \hfil 0.026\% &  \\
n$_{>3}$s REC + 82keV & \hfil 0.028\% &  \\
3p REC + 82keV & \hfil 0.036\% &  \\
3s/3p EC + 82keV & \hfil 0.027\% &  \\
3d EC + 82keV & \hfil 0.038\% &  \\
Any + 82keV & \hfil 0.26\% &  \\
    \end{tabular}
  \end{center}
\end{table}

\subsection{Acknowledgments}
We would thanks the student, G. Bertuolo that collaborated with us to the measurements on the first LYSO detector prototypes. Moreover we are grateful to UniTN Laboratory colleagues: M. Hueller, M.T. López-Arias Montenegro, P. Minati and M. Di Mauro for their support to the HP-Ge operations.

\nocite{*}

\bibliography{apssamp}

\end{document}